%% file: SecurityAPIMisuse.tex
\documentclass[sigconf]{acmart}
\setcopyright{none}
\renewcommand\footnotetextcopyrightpermission[1]{}
\usepackage{listing}
\usepackage{listings}
\usepackage{algorithmic}
\usepackage{array}
\usepackage[lined, algonl, ruled, boxed]{algorithm2e} 
\usepackage{multirow}
\usepackage{makecell}
\usepackage{xspace}
\usepackage{tcolorbox}

\newcommand{\tool}{\textsc{Seader}\xspace}
\newcommand{\todo} [1]{\textcolor{blue}{{\sf TODO}: #1}}
\newcommand{\exampleCount}{25\xspace}
\newcommand{\vulCount}{32\xspace}

\newcommand{\templateCount}{18\xspace}

\newcommand{\vcount}{988\xspace}
\newcommand{\positiveCount}{18\xspace}
\newcommand{\vulCountSecond}{39\xspace}
\newcommand{\prCount}{59\xspace}
\newcommand{\vulConfirmed}{14\xspace}
\newcommand{\codefont}[1]{\footnotesize{\texttt{#1}}\normalsize}
\newcommand{\insecure}{\textbf{\emph{I}}\xspace}
\newcommand{\secure}{\textbf{\emph{S}}\xspace}
\newcommand{\p}{\textbf{\emph{P}}\xspace}

\newcolumntype{L}[1]{>{\raggedright\let\newline\\\arraybackslash\hspace{0pt}}m{#1}}
\newcolumntype{C}[1]{>{\centering\let\newline\\\arraybackslash\hspace{0pt}}m{#1}}
\newcolumntype{R}[1]{>{\raggedleft\let\newline\\\arraybackslash\hspace{0pt}}m{#1}}

\AtBeginDocument{%
  \providecommand\BibTeX{{%
    \normalfont B\kern-0.5em{\scshape i\kern-0.25em b}\kern-0.8em\TeX}}}

\begin{document}

\title{Data-Driven Vulnerability Detection and Repair in Java Code}

\author{Ying Zhang, Mahir Kabir, Ya Xiao, Danfeng (Daphne) Yao, Na Meng}
\affiliation{%
  \institution{Computer Science, Virginia Tech}
}
\email{{yingzhang, mdmahirasefk, yax99, danfeng, nm8247}@vt.edu}

\input{abstract}



\keywords{
vulnerability detection, vulnerability repair, inter-procedural}

\maketitle
\pagestyle{plain}

\input{Introduction}
\input{Motivating_Example}

\input{approach}
\input{evaluation}

\vspace{-0.5em}
\input{Related}
\vspace{-0.5em}
\input{threats}
\vspace{-0.5em}
\input{Conclusion}

\bibliographystyle{ACM-Reference-Format}
\bibliography{reference}

\end{document}

%% file: abstract.tex
\begin{abstract}

Java platform provides various APIs to facilitate secure coding. However, correctly using security APIs is usually challenging for developers who lack cyber security training. Prior work shows that many developers misuse security APIs; such misuses can introduce vulnerabilities into software, void security protections, and present security exploits to hackers. To eliminate such API-related vulnerabilities, 
this paper presents \tool---our new approach that detects and repairs 
security API misuses. 
Given an exemplar insecure code snippet and its secure counterpart, \tool compares the snippets and conducts data dependence analysis to infer the security API misuse templates and corresponding fixing operations. Based on the inferred information, given a program, \tool performs inter-procedural static analysis to search for any security API misuse and to propose customized fixing suggestions for those vulnerabilities. 

To evaluate \tool, we applied it to \exampleCount <insecure, secure> code pairs, and \tool successfully inferred \templateCount unique API misuse templates and related fixes. With these vulnerability repair patterns, we further applied \tool to 10 open-source projects that contain in total \vulCount known vulnerabilities. 
 Our experiment shows that \tool detected vulnerabilities with 100\% precision, 84\% recall, and 91\% accuracy. 
 Additionally, we applied \tool to 100 Apache open-source projects and detected \vcount vulnerabilities; \tool always customized repair suggestions correctly. 
 Based on \tool's outputs, we filed 60 pull requests. Up till now, developers of \positiveCount projects have offered positive feedbacks on \tool's suggestions. Our results indicate that \tool can effectively help developers detect and fix security API misuses. 
Whereas prior work 
either detects API misuses \emph{or} suggests simple fixes, \tool is the first tool to do both for nontrivial vulnerability repairs.  


   
\end{abstract}

%% file: Introduction.tex
\section{Introduction}
Java platform provides libraries (e.g., Java Cryptography Architecture (JCA)~\cite{jca} and Java Secure Socket Extension~\cite{jsse}) to ease developers' secure software development (e.g., implementing key generation and secure communication). 
However, these libraries are not easy to use for two reasons. First, some APIs have overly complicated usage protocols that are poorly documented~\cite{green2016developers,Nadi:2016}. Second, developers lack the necessary cyber security training for secure coding~\cite{developer-lack-skills,meng2018secure,too-few-cybersecurity}.
Meanwhile, prior work shows that developers misused security APIs~\cite{fischer2017stack,Rahaman2019}, and  introduced vulnerabilities when building security functionalities~\cite{fahl2012eve,georgiev2012most}. For instance, Fischer et al.~found that the security API misuses posted on StackOverflow~\cite{so} were copied and pasted into 196,403 Android applications available on Google Play~\cite{fischer2017stack}. Rahaman et al.~revealed similar API misuses in 39 high-quality Apache projects~\cite{Rahaman2019}. Fahl et al.~\cite{fahl2012eve} and Georgiev et al.~\cite{georgiev2012most} separately showed that such API-related vulnerabilities could be exploited by hackers to steal data (e.g., user credentials). 

Existing tools provide insufficient support to help developers eliminate security API misuses. Specifically, some tools identify API misuses based on either hardcoded rules, rules specified with a domain-specific language, or machine learning~\cite{egele2013empirical,fischer2017stack,Li2018,kruger2018crysl,Rahaman2019,Fischer2019}. 
However, these tools cannot be easily extended to reveal new vulnerabilities, neither can they suggest any security repair. 
General-purpose program repair tools rely on automatic testing to reveal bugs, and only suggest trivial fixes (e.g., single-line edits) in very limited cases~\cite{Long2016:Prophet,Qi2014,DeMarco2014,Kim2013:PAR}. These tools cannot repair security API misuses because (1) vulnerabilities seldom fail tests, and (2) some security patches involve 
nontrivial edits like overriding an interface method. To design a better approach that overcomes the above-mentioned limitations, we need to solve two technical challenges. First, the new approach should accurately detect API misuses and suggest fixes. Second, the approach should be flexibly extensible to handle new vulnerabilities. 


\begin{figure}[h]
    \centering
    \includegraphics[width=\linewidth]{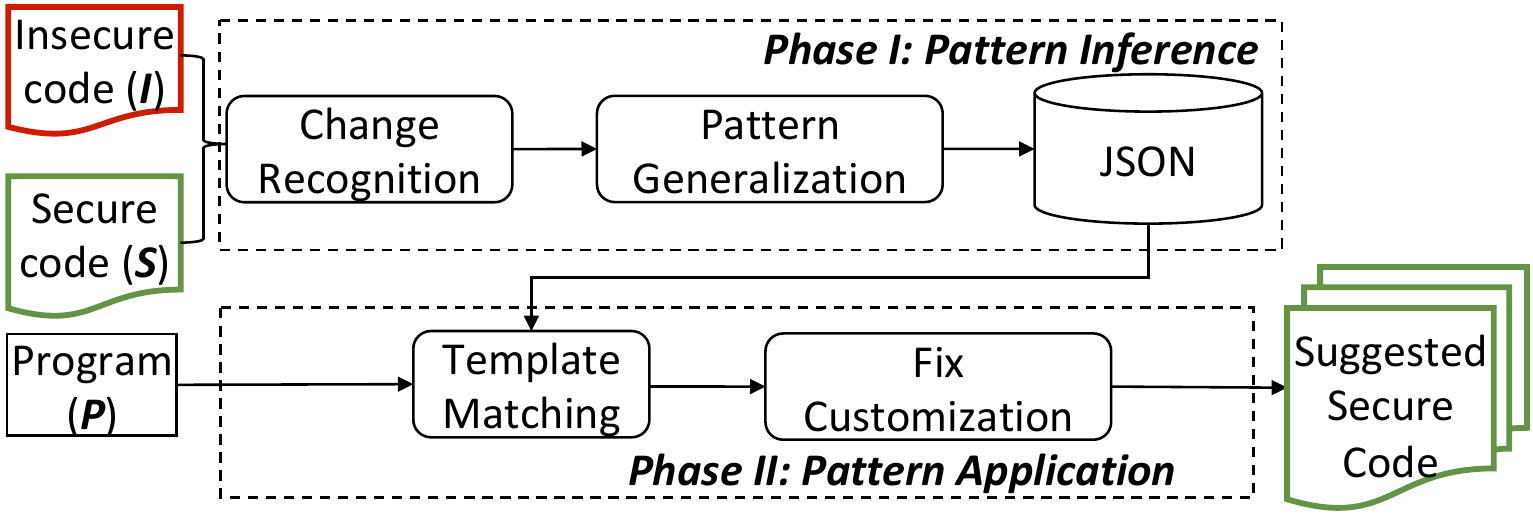}
    \vspace{-2.5em}
    \caption{The overview of \tool}
    \label{fig:overview}
    \vspace{-1.5em}
\end{figure}

In this paper, we present our new approach \tool, which detects the vulnerabilities caused by security API misuses and suggests repairs. As shown in Figure~\ref{fig:overview}, there are two phases in \tool: pattern inference and pattern application. In the first phase, suppose that a domain expert (e.g., security researcher) provides

\begin{itemize}
\item \emph{\textbf{I}}---insecure code with certain security API misuse, and
\item \emph{\textbf{S}}---the secure counterpart showing the correct API usage.
\end{itemize}
\tool compares the two code snippets and detects program changes that can transform \emph{\textbf{I}} to \emph{\textbf{S}}. Next, based on those changes, \tool derives a vulnerability repair pattern from both snippets. Each pattern  
consists of two parts: (i) a vulnerable code template together with matching-related information, and (ii) the abstract fix. \tool stores all inferred patterns into a JSON file. 
In Phase II, given a program \textbf{\emph{P}}, \tool loads all patterns from the JSON file, and searches for code matching any template. 
For each code match, 
\tool concretizes the corresponding abstract fix, and suggests code replacements  to developers. 

Additionally, security researchers have summarized a number of typical API misuse patterns related to Java security libraries~\cite{fischer2017stack,Rahaman2019}. Based on our experience with example definition for these patterns, we realized that some patterns (e.g., multiple secure options for a parameter) cannot be easily described with plain Java code examples. Therefore, we also defined an example definition library (EDL), which provides APIs 
for security experts to specify examples, and for \tool to specialize pattern inference in certain scenarios. 

For evaluation, we crafted \exampleCount <insecure, secure> code pairs based on some known API misuse patterns summarized by prior research. After \tool inferred patterns from those pairs, we further applied \tool to two program data sets to evaluate its effectiveness in terms of vulnerability detection and repair suggestion. 

Specifically, the first data set 
contains 10 Apache open-source projects, with \vulCount real vulnerabilities manually identified in the JAR files. 
With this data set, we observed \tool to detect vulnerabilities with 100\% precision, 84\% recall, and 91\% accuracy. 
The second data set contains 100 Apache open-source projects. \tool found \vcount vulnerabilities in them and proposed corresponding repair suggestions. We manually checked 100 repair suggestions and found them to be correctly customized. 
Finally, 
we filed \prCount pull requests based on \tool's reports for \prCount vulnerable Java classes and sought for developers' opinions. Up till now, the maintainers of 
\vulConfirmed vulnerable classes have confirmed the reported vulnerabilities, and 6 of them have taken actions to fix or document the vulnerabilities. 

To sum up, we made the following contributions in this paper:
\begin{itemize}
   \item We designed and implemented \tool---a data-driven approach that infers vulnerability repair patterns from <insecure, secure> code examples, and applies those patterns to detect and fix security API misuses. 
   \item For pattern inference, \tool provides a software library---EDL---for users to adopt so that they can define examples for some tricky scenarios. Our experiment with \exampleCount code pairs shows that \tool always infers patterns correctly from <insecure, secure> code examples. 
    \item In terms of vulnerability detection, our evaluation with real vulnerabilities from open-source projects shows that \tool worked as effectively as or even better than the state-of-the-art tool---CryptoGuard---in 8 of the 10 projects.  
    \item In terms of vulnerability repair, our manual inspection based on another data set shows that the repairing suggestions generated by \tool are correct. Interestingly, developers have mixed opinions on \tool's repair suggestions. 
    \end{itemize}
\tool has the data-driven nature by inferring vulnerability repair patterns from code examples, which functionality ensures the approach extensibility when new API misuses are revealed. 
\tool can accurately detect vulnerabilities and suggest meaningful fixes, showing great potential of helping eliminate API-related vulnerabilities.  

%% file: Motivating_Example.tex
\vspace{-1em}
\section{A Motivating Example}
\label{sec:motivate}
This section overviews our approach with a set of code examples. 
Prior research shows that the security of symmetric encryption schemes depends on the secrecy of any shared key~\cite{egele2013empirical}. Thus, developers should not generate secret keys from constant values hardcoded in software applications~\cite{fischer2017stack}. Suppose that 
a security expert Alex wants to detect and fix the vulnerable code that creates a secret key from a constant value. To achieve the goal with \tool, Alex needs to craft (1) an insecure code example showing the API misuse, and (2) a secure example to demonstrate the correct API usage. As shown in Figure~\ref{fig:example}, the insecure code \insecure invokes the constructor of \codefont{SecretKeySpec} by passing in a constant byte array (i.e., \codefont{StringLiterals.CONSTANTS.getBytes()}) as the first parameter. On the other hand, the secure code \secure invokes the same constructor by sending in \codefont{keyBytes} instead, which parameter is created based on a randomly generated number (see lines 1-4). 

\begin{figure}[h]
\footnotesize
\vspace{-1em}
\begin{tabular}{p{\linewidth}} \hline
\multicolumn{1}{c}{\textbf{Insecure code (\emph{I})}} \\ \hline
\vspace{-1em}
\begin{lstlisting}
SecretKey key = new SecretKeySpec(StringLiterals.CONSTANT.getBytes(), "AES");
\end{lstlisting} 
\\ \hline
\multicolumn{1}{c}{\textbf{Secure code (\emph{S})}} \\ \hline
\vspace{-1em}
\begin{lstlisting}
SecureRandom random = new SecureRandom(); 
String defaultKey = String.valueOf(random.nextInt());
byte[] keyBytes = defaultKey.getBytes(); 
keyBytes = Arrays.copyOf(keyBytes,24);
SecretKey key = new SecretKeySpec(keyBytes, "AES");
\end{lstlisting}\\ \hline
\end{tabular}
\vspace{-1.5em}
\caption{A pair of examples to show the vulnerability and repair relevant to secret key creation}\label{fig:example} 
\end{figure}
\vspace{-1em}

\begin{figure}[h]
\footnotesize
\vspace{-1em}
\begin{tabular}{p{\linewidth}} \hline
\multicolumn{1}{c}{\textbf{Vulnerable code template (T)}} \\ \hline
\vspace{-1em}
\begin{lstlisting}
SecretKey $v_0 = new SecretKeySpec(StringLiterals.CONSTANT.getBytes(), "AES");
\end{lstlisting} 
\\ \hline
\textbf{Matching-related data:}  \\
\hspace{1em}critical API: javax.crypto.spec.SecretKeySpec.SecretKeySpec(byte[], String) 

\hspace{1em}other security APIs: \{\}
\\ \hline
\multicolumn{1}{c}{\textbf{Abstract fix (F)}} \\ \hline
\vspace{-1em}
\begin{lstlisting}
SecureRandom $v_1 = new SecureRandom(); 
String $v_2 = String.valueOf($v_1.nextInt());
byte[] $v_3 = $v_2.getBytes(); 
$v_3 = Arrays.copyOf($v_3, 24);
SecretKey $v_0 = new SecretKeySpec($v_3, "AES");
\end{lstlisting}\\ \hline
\end{tabular}
\vspace{-1.5em}
\caption{The pattern inferred from the code pair in Figure~\ref{fig:example}}\label{fig:pattern} 
\end{figure}

\vspace{-0.5em}
Taking the two examples as input, \tool generates abstract syntax trees (ASTs) and compares them for any code change.
Specifically, five operations are detected: one expression update (\codefont{StringLiterals. CONSTANT.getBytes()} replaced by \codefont{keyBytes}), and four statement insertions. 
Next, based on the updated expression in \insecure, \tool conducts data dependency analysis to reveal any security API that uses the expression, and treats it as a \textbf{critical API}. 
 Such critical APIs are important for \tool to later reveal similar vulnerabilities in other codebases.
Afterwards, \tool generalizes
a \textbf{vulnerability repair pattern} from the examples by abstracting concrete variable names. 
As shown in Figure~\ref{fig:pattern}, the generalized pattern has two parts: 
(i) the vulnerability template (T) together with matching-related data and (ii) an abstract fix (F). 
Such pattern generalization ensures the inferred program transformation to be applicable
to codebases that use different variable identifiers from the given examples.

\lstset{
numbers=left, 
basicstyle=\footnotesize,
breaklines=true,
language=java,
belowskip=8pt,
escapeinside={(*}{*)},
frame = tb
}
\begin{lstlisting}[caption=A simplified version of \p,label=lst:p]
public class CEncryptor {
  private char[] passPhrase;
  private String alg = "AES";
  public CEncryptor(String passPhrase) {
    this.passPhrase = passPhrase.toCharArray();
  }
  public Result encrypt(byte[] plain) throws Exception {
    SecretKey secret = new SecretKeySpec(new String(passPhrase).getBytes(), alg);
  ...
}
public class Main {
  public static void main(String[] args)
  CEncryptor aes0 = new CEncryptor("password");
  aes0.encrypt((byte[])args[0]);
  ...
}
\end{lstlisting}

With a pattern inferred from the provided code pair, Alex can further apply \tool to an arbitrary program \p to detect and resolve any occurrence of the described vulnerability. In particular, given a program whose simplified version is shown in Listing~\ref{lst:p}, \tool first scans for any invocation of the critical API \codefont{SecretKeySpec(...)}. If no such invocation exists, \tool concludes that \p does not have the above-mentioned vulnerability; otherwise, if the API is invoked (see line 8 in Listing~\ref{lst:p}), \tool further searches for any code match for the template in Figure~\ref{fig:pattern}. Specifically, the template matching process  conducts inter-procedural analysis and checks for two conditions:

\begin{enumerate} 
\item[C1:] Does the API call have the first parameter derive from a constant?
\item[C2:] Does the API call have the second parameter match \codefont{"AES"}?
\end{enumerate}

If any invocation of \codefont{SecretKeySpec(...)} satisfies both conditions, \tool concludes that the code has the above-mentioned vulnerability. Notice that if we simply check line 8 of Listing~\ref{lst:p}, it seems that neither \codefont{new String(passPhrase).getBytes()} nor \codefont{alg} satisfies any condition. Thanks to the usage of inter-procedural analysis, \tool is able to conduct backward slicing to trace how both parameters are initialized before their current usage. With more details, because \codefont{alg} is a private field of \codefont{CEncryptor}, whose value is initialized on line 3 with \codefont{"AES"}, \tool decides that C2 is satisfied. Similarly, \codefont{passPhrase} is another field whose value is initialized with a parameter of the constructor \codefont{CEncryptor(...)} (lines 4-6). When \codefont{CEncryptor(...)} is called with parameter \codefont{"password"} before 
the invocation of \codefont{SecretKeySpec(...)} (lines 7-14), C1 is satisfied. Therefore, \tool concludes that line 8 matches the template; it thus matches concrete variable \codefont{secret} with  the template variable \codefont{\$v\_0}. 

\vspace{-0.5em}
\begin{lstlisting}[caption=A customized fix for \p suggested by \tool,label=lst:repair]
SecureRandom $v_1 = new SecureRandom(); 
String $v_2 = String.valueOf($v_1.nextInt());
byte[] $v_3 = $v_2.getBytes(); 
$v_3 = Arrays.copyOf($v_3,24);
SecretKey secret = new SecretKeySpec($v_3, "AES");
\end{lstlisting}

Based on the found code match, \tool customizes the abstract fix shown in Figure~\ref{fig:pattern} by replacing the abstract variable \codefont{\$v\_0} with concrete variable \codefont{secret}.
\tool then suggests the code snippet shown in Listing~\ref{lst:repair} to replace line 8 of Listing~\ref{lst:p}. In the suggested fix, a \codefont{SecureRandom} instance is first initialized to randomly generate a string value (lines 1-2 in Listing~\ref{lst:repair}). Next, the string is converted to a byte array (line 3), in order to match the data type of the first parameter of \codefont{SecretKeySpec(byte[], String)}. Afterwards,  the byte array is further converted to a 24-byte array (line 4), because the AES algorithm is capable of using cryptographic keys of 16, 24, or 32 bytes (i.e., 128, 192, or 256 bits)~\cite{aes}. Here, for simplicity, we randomly pick 24 bytes to create a well-formatted variable \codefont{keyBytes}. Finally, \codefont{keyBytes} and \codefont{"AES"} are passed to invoke the critical API.

%% file: Approach.tex
\vspace{-0.5em}
\section{Approach}
\label{sec:approach}

There are two phases in \tool (see Figure~\ref{fig:overview}). 
In this section, we first summarize the steps in each phase and then describe each step in detail (Section~\ref{sec:change}-Section~\ref{sec:fix}). Next, we introduce EDL---the software library we created---to facilitate example definition by security experts (Section~\ref{sec:edl}). 

\begin{description}
\item[\textbf{Phase I: Pattern Inference}]
\end{description}
\begin{itemize}
\item Given an <$I$, $S$> example pair, \tool builds an AST for each example; it then compares the ASTs to infer an edit script that can transform I to S. We denote the edit script with $E=\{op_1, op_2, \ldots, op_n\}$.
\item For any update $u\in E$ that changes an expression $e$ to $e'$, \tool analyzes data dependencies to locate the security API invoked with $e$ or $e'$, and treats the API \emph{critical}. If there is no update operation in $E$, \tool searches for any overridden security API or deleted API call. 
Next, \tool infers a vulnerability repair pattern and stores it in a JSON file. 
\end{itemize}
\begin{description}
\item[\textbf{Phase II: Pattern Application}]
\end{description}
\begin{itemize}
\item Given a program $P$, for each pattern $Pat=<T, F>$ in the JSON file, \tool searches for any code invoking the critical API. If the API is invoked, \tool further performs slicing based on the API invocation to locate other statements matching the remaining part of $T$. 
\item If $T$ is fully matched by a code snippet, \tool extracts the mappings between abstract variables and concrete variables, and then customizes the abstract fix $F$ by replacing abstract variables with concrete ones. In this way, \tool reports a vulnerability and suggests a customized fix.
\end{itemize}

\subsection{Change Recognition}
\label{sec:change}
Given an <$I$, $S$> example pair, \tool performs syntactic program differencing to identify the edit operation(s) that can change $I$ to $S$. This step consists of two parts: statement-level change recognition and expression-level change recognition. 

\begin{figure*}
\includegraphics[width=0.9\textwidth]{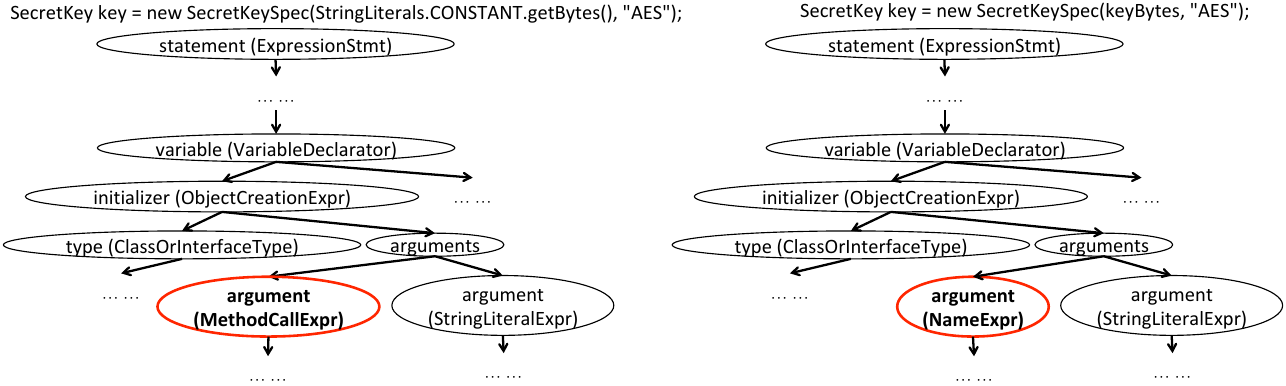}
\vspace{-1.5em}
\caption{The simplified ASTs of the two statements related to a statement-level update operation}
\label{fig:asts}
\vspace{-0.5em}
\end{figure*}

\subsubsection{Statement-level change recognition}
\tool first adopts JavaParser~\cite{hosseini2013javaparser} to generate ASTs for $I$ and $S$. Next, \tool compares statement-level AST nodes between the trees to generate an AST edit script that may contain three types of edit operations:

 \begin{itemize} 
\item \textbf{delete (Node $a$)}: Delete node $a$.

\item \textbf{insert (Node $a$, Node $b$, int $k$)}: Insert node $a$ and position it
as the $(k+1)^{th}$ child of node $b$.     

\item \textbf{update (Node $a$, Node $b$)}: Replace $a$ with
  $b$.  This operation changes $a$'s content. 
\end{itemize} 
Specifically, when comparing any two statements $s_i\in I$ and $s_s\in S$, \tool checks whether the code string of $s_i$ exactly matches that of $s_j$; if so, \tool considers $s_i$ unchanged while $I$ is transformed to $S$. Otherwise, if the code strings of $s_i$ and $s_j$ are different, \tool normalizes both statements by replacing concrete variables (e.g., \codefont{key}) with abstract variables (e.g., \codefont{\$v\_0}), and replacing constant values (e.g., \codefont{"AES"}) with abstract constants (e.g., \codefont{\$c\_0}). We denote the normalized representations as $n_i$ and $n_s$. Next, \tool computes the Levenshtein edit distance~\cite{levenshtein1966bcc} between $n_i$ and $n_s$, and computes the similarity score~\cite{fluri2007change} with: 

\vspace{-1em}
\begin{equation}
sim = 1 - \frac{edit\_distance}{max\_length(n_i, n_s)}
\end{equation}  
The similarity score $sim$ is within [0, 1]. When $sim=1$, $n_i$ and $n_j$ are identical. We set a threshold $th=0.8$ such that if   $sim>=th$, $n_i$ and $n_j$ are considered to be similar enough to match each other. In this way, \tool can detect updated statements and infer update operation(s). Compared with string-based match, the normalization-based match is more flexible, because it can match any  two statements that have similar syntactic structures but distinct variables or constants. 
Finally, if a statement $s_i\in I$ does not find a match in $S$, \tool infers a delete operation; if $s_s\in S$ is unmatched, \tool infers an insert operation.  

\vspace{-0.5em}
\subsubsection{Expression-level change recognition}
When $s_i$ is updated to $s_s$, there is usually a portion of $s_i$ changed. The finer-granularity edit in the statement (e.g., expression replacement) can help us better interpret changes and conduct  pattern generalization (see Section~\ref{sec:generalize}). Therefore, for each statement-level update, \tool further identifies any finer-granularity edit by conducting
top-down matching between the ASTs of $s_i$ and $s_s$. Specifically, 
while traversing both trees in a preorder manner, \tool compares roots and inner nodes based on the AST node types and compares leaf nodes based on the code content. Such node traversal and comparison continue until all unmatched subtrees or leaf nodes are found. 

In particular, for the example shown in Section~\ref{sec:motivate}, after \tool performs statement-level change recognition, it reveals one statement update together with four statement insertions. Figure~\ref{fig:asts} illustrates the simplified ASTs of the statements involved in the update. By comparing the nodes of both ASTs in a top-down manner, \tool detects that the first arguments sent to the constructor function differ (e.g., \codefont{MethodCallExpr} vs.~\codefont{NameExpr}). Therefore, \tool creates a finer-granularity operation to replace the statement-level update: \textbf{update (\codefont{StringLiterals.CONSTANT.getBytes()}, \codefont{keyBytes})}. 
 
 \vspace{-0.5em}
\subsection{Pattern Generalization}
\label{sec:generalize}
When security experts present an <$I$, $S$> example pair to demonstrate any API misuse, we expect that they provide the minimum code snippets to show only one vulnerability and its repair. Additionally, based on our experience with security API misuses, each vulnerability is usually caused by the misuse of one security API. Therefore, to infer a general vulnerability-repair pattern, we always analyze the given examples to answer two questions: 

\begin{enumerate}
\item[1.] What is the security API whose misuse is responsible for the vulnerability (i.e., critical API)? 
\item[2.] What is the relationship between the critical API and its surrounding code?
\end{enumerate}

\subsubsection{Task 1: Identifying the critical API} Starting with the edit script $E$ created in Section~\ref{sec:change}, \tool looks for any update operation $update(e, e')$. If there is such an operation, \tool searches for the security API whose invocation is data-dependent on $e$ or $e'$, and considers the API to be \emph{critical}. For the example shown in Figure~\ref{fig:asts}, the critical API is \codefont{SecretKeySpec(byte[], String)} because it is invoked with the updated expression as the first argument.
Similarly, Figure~\ref{fig:example2} presents another example where a numeric literal is updated from \codefont{4} to \codefont{8}. With data dependency analysis, \tool reveals that the constants are used to define variable \codefont{salt}, while \codefont{salt} is used as an argument when \codefont{PBEParameterSpec(...)} is invoked. Therefore, the method invocation depends on the updated expression, and the security API \codefont{PBEParameterSpec(byte[], int)} is considered \emph{critical}.

\lstset{
numbers=left, 
basicstyle=\footnotesize,
breaklines=true,
language=java,
belowskip=-5pt,
escapeinside={(*}{*)},
frame=none
}

\begin{figure}
\footnotesize
\begin{tabular}{p{\linewidth}} \hline
\multicolumn{1}{c}{\textbf{Insecure code (\emph{I})}} \\ \hline
\vspace{-1em}
\begin{lstlisting}
void test(int iterations) {
  byte[] salt = new byte[4];
  AlgorithmParameterSpec paramSpec = new PBEParameterSpec(salt, iterations);
}
\end{lstlisting} 
\\ \hline
\multicolumn{1}{c}{\textbf{Secure code (\emph{S})}} \\ \hline
\vspace{-1em}
\begin{lstlisting}
void test(int iterations) {
  byte[] salt = new byte[8];
  AlgorithmParameterSpec paramSpec = new PBEParameterSpec(salt, iterations);
}
\end{lstlisting}\\ \hline
\end{tabular}
\vspace{-1.5em}
\caption{A pair of examples where the updated constant is indirectly depended on by the invocation of critical API \codefont{PBEParameterSpec(byte[], int)}}\label{fig:example2} 
\vspace{-2em}
\end{figure}

If there is no update operation in $E$, \tool searches for any overridden security API that encloses all edit operations, and considers the overridden API to be \emph{critical}. Take the code pair shown in Figure~\ref{fig:example3} as an example. By comparing $I$ with $S$, \tool can identify one statement deletion and multiple statement insertions. As there is no update operation and all edit operations are enclosed by an overridden method \codefont{verify(String, SSLSession)} (indicated by \codefont{@Override}), \tool further locates the interface or super class declaring the method (e.g., \codefont{HostnameVerifier}). If the overridden method together with the interface/super class matches any known security API, \tool concludes the overridden method to be a \emph{critical API}. 

Lastly, if no update operation or overridden security API is identified, \tool checks whether there is any deletion of security API call in $E$; if so, the API is \emph{critical}. 
To facilitate later template matching (Section~\ref{sec:match}), for each identified critical API, \tool records the method binding information (e.g., \codefont{javax.crypto.spec.SecretKeySpec. SecretKeySpec(byte[], String)}). 

\begin{figure}
\footnotesize
\begin{tabular}{p{\linewidth}} \hline
\multicolumn{1}{c}{\textbf{Insecure code (\emph{I})}} \\ \hline
\vspace{-1em}
\begin{lstlisting}
public class HostVerifier implements HostnameVerifier {
  @Override
  public boolean verify(String hostname, SSLSession sslSession){
    return true;
  }}
\end{lstlisting} 
\\ \hline
\multicolumn{1}{c}{\textbf{Secure code (\emph{S})}} \\ \hline
\vspace{-1em}
\begin{lstlisting}
public class HostVerifier implements HostnameVerifier {
  @Override
  public boolean verify(String hostname, SSLSession sslSession){
    //Please change "example.com" as needed
    if ("example.com".equals(hostname)) { 
      return true;
    }
    HostnameVerifier hv = HttpsURLConnection.getDefaultHostnameVerifier(); 
    return hv.verify(hostname, sslSession);  
  }}
\end{lstlisting}\\ \hline
\end{tabular}
\vspace{-1.5em}
\caption{A pair of examples from which \tool infers the critical API to be an overridden method}\label{fig:example3} 
\vspace{-2em}
\end{figure}

\subsubsection{Task 2: Extracting relationship between the critical API and its surrounding code}
When a vulnerable code example contains multiple statements (e.g., Figure~\ref{fig:example2} and Figure~\ref{fig:example3}), we were curious how the critical API invocation is related to other statements. 
On one extreme, if the invocation is irrelevant to all surrounding statements, we should not include any surrounding code into the generalized pattern. On the other extreme, if the invocation is related to all surrounding code, we should take all code into account when inferring a vulnerability-repair pattern. Thus, this task intends to decide (1) which statements of $I$ to include into the vulnerable code template, (2) what other security API(s) whose invocations should be analyzed for template matching (see Section~\ref{sec:match}), and (3) which statements of $S$ to include into the abstract fix.


Specifically, \tool performs intra-procedural data dependency analysis. If any statement defines a variable value that is (in)directly used by the critical API invocation, the statement is extracted as \emph{edit-relevant context}. \tool relies on such context to characterize the demonstrated vulnerability. 
For the insecure code $I$ shown in Figure~\ref{fig:example2}, because the API call (line 3) data-depends on variable \codefont{salt}, lines 2-3 are extracted as the context. Additionally, when the critical API is an overridden method, its code implementation in $I$ is considered as edit-relevant context (see lines 3-5 in Figure~\ref{fig:example3}). Based on the extracted edit-relevant context, \tool abstracts all variables to derive a vulnerable code template $T$, and records mappings $M$ between abstract and concrete variables. 

In addition to the critical API, \tool also extracts binding information for any other security API invoked by the context code. Compared with edit-relevant context, these APIs provide more succinct hints of code vulnerabilities. In our later template matching process, these APIs can serve as ``\emph{anchors}'' for \tool to efficiently decide whether a program slice invokes all relevant APIs or is worth further comparison with the template $T$. 

To determine the fix-relevant code in secure version $S$, \tool identifies any unchanged code in the edit-relevant context, the inserted statements, and the new version of any updated statement. For the secure code $S$ shown in Figure~\ref{fig:example2}, lines 2-3 are fix-relevant, because line 2 is the new version of an updated statement and line 3 is unchanged context code. Similarly, for the secure code $S$ shown in Figure~\ref{fig:example3}, lines 3-10 are fix-relevant, because lines 3 and 10 present the critical API while lines 4-9 are inserted statements. Based on the above-mentioned variable mappings $M$ and fix-related code, \tool further abstracts variables used in the fix-related code to derive an abstract fix $F$. \tool ensures that the same concrete variables used in $I$ and $S$ are always consistently mapped to the same abstract variables recorded in $M$. 

To sum up, for each given <$I$, $S$> pair, \tool produces a pattern $Pat=<T, F>$, which consists of a vulnerable code template $T$, an abstract fix $F$, and metadata to describe $T$ (i.e., bindings of the critical API and other invoked security APIs). 

\input{algorithm}

\vspace{-0.5em}
\subsection{Template Matching}
\label{sec:match}
Given a program $P$, \tool uses a static analysis framework---WALA~\cite{wala}---to analyze the JAR file (i.e., byte code) of $P$. 
As shown in Algorithm~\ref{alg1}, to find any code in $P$ that matches the template $T$, \tool first searches for the critical API (i.e., invocation or method reimplementation). If the critical API does not exist, \tool concludes that there is no match for $T$. Next, if the critical API is invoked at least once, for each invocation, \tool conducts inter-procedural backward slicing to retrieve all code $Sli$ on which the API call is either control- or data- dependent (i.e., \codefont{getBackwardSlice(x)}). When $T$ invokes one or more security APIs in addition to the critical API, \tool further examines whether $Sli$ contains matches for those extra APIs; if not, the matching trial fails. Next, \tool checks whether the matched code in $Sli$ preserves the data dependencies manifested by $T$ (i.e., \codefont{dataDependConsist(T, Sli}).  
If those data dependencies also match, \tool reveals a vulnerability.  

Alternatively, if the critical API is reimplemented, for each reimplementation, \tool compares the code content against $T$, and reports a vulnerability if they match. At the end of this step, if any vulnerability is detected, \tool presents the line number where the critical API is invoked or is declared as an overridden method, and shows related matching details. The matching details include both code matches and abstract-concrete variable mappings. 

\vspace{-0.5em}
\subsection{Fix Customization}
\label{sec:fix}
This step involves two types of customization: variable customization and edit customization. To customize variables, based on the matching details mentioned in Section~\ref{sec:match}, \tool replaces abstract variables in $F$ with the corresponding concrete ones. 
We denote this customized version as $F_c$. For edit customization, \tool suggests code replacements in two distinct ways depeding on the inferred edit operations mentioned in Section~\ref{sec:change}. Specifically, if there is only one update operation inferred, \tool simply recommends an alternative expression to replace the original expression. Otherwise, \tool presents $F_c$ for developers to consider.

Notice that \tool does not directly modify $P$ to automatically repair any vulnerability for two reasons. First, when template $T$ contains multiple statements, it is possible that the corresponding code match involves statements from multiple method bodies. Automatically editing those statements can be risky and can unpredictably impact  program semantics. 
Second, some fixes require for developers' further customization  based on their software environments (e.g., network configurations, file systems, and security infrastructures). As implied by Figure~\ref{fig:example3}, the abstract fix derived from $S$ will contain a comment \codefont{"//Please change 'example.com' as needed"}, so will the customized fix by \tool. This comment instructs developers to replace the standard hostname based on their needs. 


\vspace{-0.5em}
\subsection{Specialized Handling for Certain Patterns}
\label{sec:edl}
We believe that by crafting <$I$, $S$> code pairs, security researchers can intuitively demonstrate the misuse and correct usage of security APIs. However, we also noticed some scenarios where plain Java examples cannot effectively indicate the vulnerability-repair patterns. To solve this problem, we defined a Java library named EDL for user adoption and invented special ways of example definition. 
Currently, EDL has two classes: \codefont{StringLiterals} and \codefont{IntLiterals}, which provide APIs to help specify constant-related examples. With the library support, \tool allows users to specially define examples for three typical scenarios: 

\paragraph{Scenario 1: A pattern involves the usage of \textbf{any} constant value instead of a \textbf{particular} constant.} Plain examples only show the usage of particular constant values, but cannot generally represent the constant concept. Consider the vulnerability introduced in Section~\ref{sec:motivate}. Without using  \codefont{StringLiterals.CONSTANT}, a domain expert  
has to define a plain example to show the API misuse, such as 

\codefont{SecretKey key = new SecretKeySpec("ABCDE".getBytes(), "AES");}

\noindent
\tool is designed to preserve all string literals from $I$ when generalizing template $T$, and to look for those values when matching code with $T$. Consequently, given the above-mentioned example, \tool will inevitably embed \codefont{"ABCDE"} into the inferred template. 
To help users avoid such unwanted literal values in $T$, EDL provides 
\codefont{StringLiterals.CONSTANT} and \codefont{IntLiterals.CONSTANT}. These APIs can be used as placeholders for string and integer constants to represent any arbitrary value in examples. 
When \tool identifies such special EDL APIs in examples, it keeps the APIs in $T$ and later specially matches them with constants in $P$.

\paragraph{Scenario 2: A pattern has multiple alternative insecure (or secure) options} Given a parameter of certain security API, suppose that there are (1) $m$ distinct values to cause API misuse and (2) $n$ other alternatives to ensure security, where $m\ge1$, $n\ge1$. To enumerate all possible combinations between the vulnerable code templates and repair options, users have to provide $m\times n$ pairs of plain examples, which practice is inefficient. To solve this issue, we defined another two method APIs in \codefont{StringLiterals}. As shown in Figure~\ref{fig:multi}, one API is a constructor of \codefont{StringLiterals}, which can take in any number of string literals as arguments (see line 1 in $I$) and store those values into an internal list structure. The other API is \codefont{getAString()}, which randomly picks and returns a value from that list (see line 2 in $I$). With these two APIs, a domain expert can efficiently illustrate multiple secure/insecure options in just one code pair. 

\begin{figure}[h]
\footnotesize
\begin{tabular}{p{\linewidth}} \hline
\multicolumn{1}{c}{\textbf{Insecure code (\emph{I})}} \\ \hline
\vspace{-1em}
\begin{lstlisting}
StringLiterals literals=new StringLiterals("AES", "RC2", "RC4", "RC5", "DES", "blowfish", "DESede");
Cipher.getInstance(literals.getAString());
\end{lstlisting} 
\\ \hline
\multicolumn{1}{c}{\textbf{Secure code (\emph{S})}} \\ \hline
\vspace{-1em}
\begin{lstlisting}
StringLiterals literals = new StringLiterals("AES/GCM/PKCS5Padding","RSA", "ECIES);
Cipher.getInstance(literals.getAString());
\end{lstlisting}\\ \hline
\end{tabular}
\vspace{-1.5em}
\caption{A code pair where multiple alternative secure and insecure options are specified simultaneously}\label{fig:multi} 
\vspace{-1em}
\end{figure}

The examples in Figure~\ref{fig:multi} show that when security API
 \codefont{Cipher. getInstance(...)} is called, the insecure parameter values include \codefont{"AES"}, \codefont{"RC2"}, \codefont{"RC4"}, \codefont{"RC5"}, \codefont{"DES"}, \codefont{"blowfish"}, and \codefont{"DESede"}. Each insecure value should be replaced by either  \codefont{"AES/GCM/PKCS5Padding"}, \codefont{"RSA"}, or \codefont{"ECIES"}. 
 Given such examples, \tool extracts insecure/secure options from \codefont{StringLiterals}-related statements, detects vulnerabilities in $P$ if the security API is invoked with any insecure option, and suggests fixes by randomly picking one of the secure options.  

\paragraph{Scenario 3: A pattern requires for certain value range of a variable.} Given an integer parameter $p$ of certain security API, suppose that there is a threshold value $th$ such that the API invocation is secure only when $p\ge th$. To enumerate all possible vulnerable cases and related repairs via plain examples, theoretically, a user should provide $(th-Integer.MIN\_VALUE) \times (Integer.MAX\_VALUE-th+1)$ code pairs, which practice is cumbersome and unrealistic. Therefore, we invented a special way of example definition for such scenarios, which requires users to provide only (1) one insecure example by setting $p$ to a concrete value less than $th$ and (2) one secure example by setting $p=th$. 
As shown in Figure~\ref{fig:example2}, if a security expert wants to describe the pattern that \emph{the array size of the first parameter used in \codefont{PBEparameterSpec(byte[], int)} should be no less than 8}, then he/she can define $I$ by creating an array with a smaller size (i.e., 4) and define $S$ by setting the array size to 8. 
\tool was developed to identify the usage of distinct integer literals between $I$ and $S$, and to infer a secure value range accordingly. 


%% file: algorithm.tex
\SetKwData{CandidateMatches}{Candi} 
\SetKwData{Matched}{Matched} 
\SetKwData{LeafMatches}{L} 
\SetKwData{SymbolicNameMapping}{S}

\begin{algorithm}
\caption{Matching Program P to template T} 
\label{alg1} 

\small
\KwIn{P, T, D \tcc{program, template, and related metadata}} 
\KwOut{Matched \tcc{a set of code matches from P to T}}

Candi $:=$ $\emptyset$, Matched $:= \emptyset$; \\
\tcc{1. search for matches of the critical API} 
\ForEach{code line x $\in$ P}{
  \If{x invokes D(critical) || x declares D(critical)}{ 
    Candi := Candi $\cup$ x; 	
  } 
}

\ForEach{x $\in $Candi}{
  \If{x invokes {\tt D}(critical)}{
\tcc{2(a). For API call, do program slicing and look for matches of other security APIs} 
Sli = getBackwardSlice(x);\\
\If{(Sli has all matches for D(other)) == false}{
continue;  
}
\tcc{3. check whether the data dependencies between security APIs in T match those in Sli}
\If{dataDependConsist(T, Sli)}{
Matched:=Matched $\cup$ \{Sli, mappings\};
}
}\Else{
\tcc{2(b). For API overriding, check the code}
\If{contentMatch(code(P, x), T)}{
Matched := Matched $\cup$ \{code(P, x), mappings\};
}
}
}
\end{algorithm}

%% file: Evaluation.tex
\vspace{-0.5em}
\section{Evaluation}

This section first introduces the data sets and metrics used for evaluation (Sections~\ref{sec:datasets}-\ref{sec:metrics}). It then presents \tool's effectiveness of pattern inference (Section~\ref{sec:infer}). Next, it describes \tool's capabilities of vulnerability detection (Section~\ref{sec:detect}) and repair suggestion (Section~\ref{sec:repair}). Essentially, Sections~\ref{sec:detect} and~\ref{sec:repair} reflect \tool's effectiveness of pattern application.

\vspace{-0.5em}
\input{Result/datasets}
\vspace{-0.5em}
\input{Result/metrics}
\vspace{-0.5em}
\input{Result/infer}
\vspace{-1.5em}
\input{Result/detect}

\vspace{-0.5em}
\input{Result/repair}

%% file: Result/datasets.tex
\subsection{Data Sets}
\label{sec:datasets}

There are two types of data used: data to evaluate pattern inference, and data to evaluate pattern application.

\begin{table*}
\centering
\caption{The API misuses and related fixes summarized by prior work~\cite{Mendel2013,tls1.2,fischer2017stack,Rahaman2019}}\label{tab:classes}
\vspace{-1em}
\footnotesize
\begin{tabular}{R{0.4cm}|p{2.5cm}|p{7cm}|p{6.5cm}}
\toprule
\textbf{Idx} & \textbf{Security Class API} & \textbf{Insecure} & \textbf{Secure}\\ \toprule
1 & \codefont{Cipher} & The algorithm and/or mode is set as RC2, RC4, RC5, DES, DESede, AES/ECB, or Blowfish. & The algorithm and/or mode is set as RSA, GCM, AES, or ECIES \\ \hline 
2 & \codefont{HostnameVerifier} & Allow all hostnames. & Implement logic to actually verify hostnames.  \\ \hline
3 & \codefont{IvParameterSpec} & Create an initialization vector (IV) with a constant. & Create an IV with a random value. \\ \hline
4 & \codefont{KeyPairGenerator} & Create an RSA key pair where key size < 2048 bits or create an ECC key pair where key size < 224 bits & RSA key size >= 2048 bits, ECC key size >= 224 bits \\ \hline
5 & \codefont{KeyStore} & When loading a keystore from a given input stream, the provided password is a hardcoded constant non-null value. & The password is retrieved from some external source (e.g., database or file)  \\ \hline
6 & \codefont{MessageDigest} & The algorithm is MD2, MD5, SHA-1, or SHA-256. & The algorithm is SHA-512.\\ \hline
7 & \codefont{PBEKeySpec} & Create a PBEKey based on a constant salt. & Use a randomly generated salt value to create the key.\\ \hline
8 & \codefont{PBEParameterSpec} & Create a parameter set for password-based encryption (PBE) by setting salt size < 64 bits or iteration count < 1000 & Salt size >= 64 bits,  iteration count >=1000. \\ \hline
9 & \codefont{SecretKeyFactory} & Create secret keys with algorithm DES, DESede, ARCFOUR, PBEWithMD5AndDES, or PBKDF2WithHmacSHA1. & Create secret keys with AES or PBEWithHmacSHA256AndAES\_128.\\ \hline
10 & \codefont{SecretKeySpec} & Create a secret key with a constant value. & Create a secret key with a randomly generated value. \\ \hline
11 & \codefont{SecureRandom} & Use Random to generate random values, or set SecureRandom to use a constant seed. & Use SecureRandom instead of Random, and ensure the seed to be a random value. \\ \hline
12 & \codefont{SSLContext} & Protocol version < TLSv1.3 & Protocol version >= TLSv1.3\\ \hline
13 & \codefont{TrustManager} & Trust all clients or servers & Check clients and/or check servers. \\ \bottomrule
\end{tabular}
\end{table*}

\subsubsection{A data set to evaluate pattern inference}
\label{sec:evaluateInfer}

Prior research summarized a number of security API misuses and related correct usage in Java~\cite{rfc7525,rfc2898,Manger2001,fahl2012eve,Mendel2013,tls1.2,fischer2017stack,meng2018secure,chen2019reliable,Rahaman2019}. To evaluate \tool's effectiveness of pattern inference, we referred to some well-described API misuses and fixes to craft code examples as inputs for \tool. With more details, Table~\ref{tab:classes} lists the 13 security class APIs we focused on, the insecure usage of certain method API(s) defined by these classes, and the related secure usage. With this domain knowledge, we manually created a set of \exampleCount <insecure, secure> code pairs.  

\subsubsection{Two data sets to evaluate pattern application}
\label{sec:evaluateApply} 



The first data set contains \vulCount real vulnerabilities from 10 Apache open-source projects. To create this set, the second author randomly picked 10 Apache projects and searched for any security API usage based on keywords. For each invoked or overridden security API, the second author leveraged his domain knowledge to manually analyze the program context and to decide whether the API is misused. Once the data set was created, another two authors checked the data to decide whether all included vulnerabilities were true positives; if any false positive was revealed, all three authors discussed until reaching an agreement. Notice that none of these three authors has experience with \tool. Because the data set was created in a tool-agnostic way, we can use it to objectively evaluate \tool's capability of vulnerability detection. We will also use this data to evaluate the effectiveness of the state-of-the-art vulnerability detection tool---CryptoGuard~\cite{Rahaman2019}, and compare \tool with CryptoGuard. 

The other data set contains 100 widely used Apache open-source projects. We decided to download Apache projects for three reasons. First, they are well-maintained and have good code quality; it can be hard for us to identify vulnerabilities in these projects. Second, the project developers are usually responsive to the pull requests or code changes suggested by other developers. Third, Apache projects are usually widely used, so any API misuses found in those projects are important and may negatively impact many other projects that depend on them. Therefore, to create this data set, we first ranked the Apache projects available on GitHub~\cite{gh} in the descending order of their stars. Namely, the more stars a project $Prj$ has, the more likely that $Prj$ is popular, and the higher $Prj$ is ranked. Next, we located the top 100 projects that meet the following two criteria: 

\begin{itemize}
\item Each project uses at least one security API that \tool examines.  
\item The project is compilable because \tool analyzes the compiled JAR files to reveal vulnerabilities. 
\end{itemize}
This data set is used to mainly evaluate \tool's effectiveness of repair suggestion. 

%% file: Result/metrics.tex
\subsection{Metrics}
\label{sec:metrics}

As with prior work~\cite{Rahaman2019}, we leveraged the following three metrics to measure tools' capability of vulnerability detection: 

\emph{\textbf{Precision (P)}} measures among all reported vulnerabilities, how many of them are true vulnerabilities. 

\vspace{-0.5em}
\begin{equation}
P=\frac{\text{\# of correct reports}}{\text{Total \# of reports}} \times 100\%
\end{equation}
\vspace{-0.5em}

\emph{\textbf{Recall (R)}} measures among all known vulnerabilities, how many of them are detected by a tool. 

\vspace{-0.5em}
\begin{equation}
R=\frac{\text{\# of correct reports}}{\text{Total \# of known vulnerabilities}} \times 100\%
\end{equation}
\vspace{-0.5em}

\emph{\textbf{F-score (F)}} is the harmonic mean of precision and recall, and is used to measure the overall detection accuracy of an approach:
\vspace{-0.5em}
\begin{equation}
F = \frac{2 \times P \times R}{P + R} \times 100\%
\end{equation}
\vspace{-1em}

%% file: Result/infer.tex
\subsection{Effectiveness of Pattern Inference}
\label{sec:infer}

\begin{table*}
\centering
\footnotesize
\caption{Evaluation results on the \vulCount-vulnerability set}\label{tab:32}
\vspace{-1.5em}
\begin{tabular}{C{4.5cm}|R{1cm}|R{1.5cm}R{1cm}R{0.5cm}R{0.5cm}R{0.5cm}|R{1.5cm}R{1cm}R{0.5cm}R{0.5cm}R{0.5cm}} \toprule
\multirow{2}{*}{\textbf{Apache Project}} & \textbf{Manual}  & \multicolumn{5}{c|}{\textbf{\tool}} & \multicolumn{5}{c}{\textbf{CryptoGuard}} \\ \cline{3-12}
& \textbf{Ground Truth (GT)}& \textbf{\# of Reported Vulnerabilities} & \textbf{\# of Correct Reports} & \textbf{P(\%)} & \textbf{R{\%}} & \textbf{F(\%)} & \textbf{\# of Reported Vulnerabilities} & \textbf{\# of Correct Reports} & \textbf{P(\%)} & \textbf{R{\%}} & \textbf{F(\%)} \\ \toprule
apacheds-kerberos-codec-2.0.0.AM25.jar & 10 & 12 & 12 & 100 & 100 & 100 & 12 & 12 & 100 & 100 & 100\\ \hline
artemis-commons-2.11.0.jar & 3 & 4 & 4 & 100 & 100 & 100 & 4 & 4 & 100 & 100 & 100 \\ \hline 
deltaspike-core-impl-1.9.2.jar & 1 & 2 & 2 & 100 & 100 & 100 & 2 & 2 & 100 & 100& 100\\ \hline
flume-file-channel-1.9.0.jar & 1 & 3 & 3 & 100 & 100 & 100 & 3 & 3 & 100 & 100 & 100\\ \hline
hadoop-hdfs-3.2.1.jar & 5 & 5 & 5 & 100 & 100 & 100 & 5 & 5 & 100 & 100 & 100\\ \hline
jclouds-core-2.2.0.jar & 2 & 7 & 7 & 100& 100& 100 & 5 & 5 & 100 & 100 & 100 \\ \hline
shiro-crypto-cipher-1.5.0.jar & 0 & 0 & 0 & 100& -& -& 0 &0&100 &- &- \\ \hline
wagon-provider-api-3.3.4.jar & 0 & 1 &1 &100 &- &- & 1 & 1& 100& -& -\\ \hline
wss4j-ws-security-common-2.2.4.jar & 5 & 5 & 5 & 100 & 100 & 100 & 5 & 5 & 100 & 100 & 100 \\ \hline
wss4j-ws-security-stax-2.2.4.jar & 5 & 0 & 0 & - & - & - & 5 & 5 & 100 & 100 & 100 \\ \bottomrule
\textbf{Overall} & 32 & \vulCountSecond & \vulCountSecond & 100 & 84 & 91 & 42 & 42 & 100 & 100 & 100 \\ \bottomrule
\multicolumn{12}{l}{When \textbf{\# of Reported Vulnerabilities} > \textbf{Manual Ground Truth}, we manually checked each additional report to decide the correctness and calculated precision accordingly.}
\end{tabular}
\vspace{-1em}
\end{table*}

As mentioned in Section~\ref{sec:datasets}, we crafted \exampleCount code pairs to evaluate \tool's effectiveness of pattern inference. To facilitate explanation, we categorized these pairs by checking two conditions: 
\begin{enumerate}
\item[C1.] Do $I$ and $S$ contain single or multiple statements? 
\item[C2.] Does the pattern generalization involve variable abstraction? 
\end{enumerate}
The two conditions actually reflect the difficulty levels or challenges of these pattern inference tasks. For instance, if $I$ or $S$ has multiple statements, \tool conducts data-dependency analysis to locate the edit-relevant context in $I$ or to reveal the fix-relevant code in $S$. If $I$ or $S$ uses variables, \tool abstracts all variable names to ensure the general applicability of inferred patterns. As shown in Table~\ref{tab:examplesInfer}, there are five simplest pairs that can be handled by \tool without conducting any data-dependency analysis or identifier generalization. Meanwhile, there are 14 most complicated cases that require for both data-dependency analysis and identifier generalization. 

\begin{table}
\caption{The \exampleCount code pairs for pattern inference}\label{tab:examplesInfer}
\vspace{-1.2em}
\footnotesize
\centering
\begin{tabular}{R{1cm}|R{2cm}|R{2.5cm}} \toprule
& \textbf{Single statement} & \textbf{Multiple statements} \\ \hline
\textbf{Identical} & 5 & 5 \\ \hline
\textbf{Abstract} & 1 & 14 \\ 
 \bottomrule
\end{tabular}\vspace{-1.5em}
\end{table}

In our evaluation, \tool correctly inferred patterns from all pairs. When some pairs illustrate distinct secure/insecure options (e.g., distinct string literals) for the same critical API, \tool merged the inferred patterns because they demonstrate misuses of the same security API. In this way, \tool derived \templateCount unique patterns. 

\noindent\begin{tabular}{|p{8cm}|}
	\hline
	\textbf{Finding 1:} \emph{Our experiment with \exampleCount  code pairs shows \tool's great capability of pattern inference; it also indicates the great potential of example-based vulnerability detection and repair.   }
\\
	\hline
\end{tabular}
\vspace{1em}

%% file: Result/detect.tex
\subsection{Effectiveness of Vulnerability Detection}
\label{sec:detect}

As described in Section~\ref{sec:evaluateInfer}, the \vulCount vulnerabilities were manually identified in 10 randomly selected Apache projects. In particular, as shown in Table~\ref{tab:32}, the manual ground truth set (GT) actually includes vulnerable code from eight projects, but covers no vulnerability in two projects: shiro-crypto-cipher-1.5.0.jar and wagon-provider-api-3.3.4.jar; thus, we were unable to calculate tools' recall rates based on GT for the two projects. 
\tool and CryptoGuard separately revealed vulnerabilities in eight and nine projects. 


For six projects (i.e., apacheds-kerberos-codec-2.0.0.AM25.jar, artemis-commons-2.11.0.jar, deltaspike-core-impl-1.9.2.jar, 
flume-file-channel-1.9.0.jar, jclouds-core-2.2.0.jar, and wagon-provider-api-3.3.4.jar), both tools revealed more vulnerabilities than GT. Because GT can be incomplete, it may miss some true vulnerabilities that either tool can identify. To properly evaluate the precision rates in such scenarios, we first intersected the tool-generated reports with GT, and then manually examined any unmatched report for its correctness. Take artemis-commons-2.11.0.jar as an example. Among the four vulnerabilities reported by \tool, three of them match the known ones in GT; thus, we further checked the remaining one and confirmed the vulnerability. In this way, the precision of \tool for the project is 4/4$\times$100\%=100\%. We observed that all reports generated by \tool and CryptoGuard are true positives, so both tools achieved 100\% precision. 

For project jclouds-core-2.2.0.jar, \tool identified more vulnerabilities than CryptoGuard (7 vs.~5), mainly because \tool checks for more insecure options of the security API \codefont{SSLContext.getInstance( String protocol)}. Meanwhile, CryptoGuard identified more vulnerabilities in the project wss4j-ws-security-stax-2.2.4.jar than \tool (5 vs.~0). This is mainly due to the limitation of WALA---the static analysis framework used by \tool. When WALA builds a call graph for the whole program $P$, it requires that all library dependencies of $P$ should exist as JAR files in the class path. If some dependencies are missing, the call graph built by WALA can be incomplete. Consequently, when \tool traversed all methods in an incomplete graph $CG$, it missed the API misuses of the methods that are excluded from $CG$.  

Overall, given 10 open-source projects, \tool reported \vulCountSecond vulnerabilities and achieved 100\% precision, 84\% recall, and 91\% accuracy. CryptoGuard detected 42 vulnerabilities and achieved 100\% precision, 100\% recall, and 100\% accuracy. The comparison shows that \tool effectively detected API-related vulnerabilities, and has comparable performance with CryptoGuard in most projects. 
In the future, we will further improve \tool by expanding its pattern set and overcoming the limitation due to WALA. 

\vspace{0.5em}
\noindent\begin{tabular}{|p{8cm}|}
	\hline
	\textbf{Finding 2:} \emph{Given 10 open-source projects, \tool worked as effectively as the state-of-the-art tool CryptoGuard in 8 projects. This indicates \tool's great capability of vulnerability detection. 
	}
\\
	\hline
\end{tabular}
\vspace{1em}

%% file: Result/repair.tex
\subsection{Effectiveness of Repair Suggestion}
\label{sec:repair}

\begin{table}
\footnotesize
    \centering
    \caption{The \vcount vulnerabilities found in 100 Apache projects}
    \label{tab:100 projects}
    \vspace{-1.5em}
    \begin{tabular}{l|r|r}
    \toprule
           \textbf{Security Class API}& \textbf{\# of Detected Vulnerabilities} & \textbf{\# of Suggested Fixes} \\ 
        \hline
        \codefont{Cipher} & 23 & 23 \\
        \hline
        \codefont{HostnameVerifier} & 68 & 68 \\ 
        \hline
        \codefont{IvParameterSpec} & 6 & 6\\
        \hline
        \codefont{KeyPairGenerator} & 1 & 1\\
        \hline
        \codefont{KeyStore} & 10 & 10\\
        \hline
        \codefont{MessageDigest} & 147 & 147 \\
        \hline
        \codefont{PBEKeySpec} &8 & 8 \\
        \hline
        \codefont{PBEParameterSpec} &6 & 6\\
        \hline
        \codefont{SecretKeyFactory} & 12 & 12\\
        \hline
        \codefont{SecretKeySpec} & 11& 11 \\
        \hline
        \codefont{SecureRandom}&422 &422\\
        \hline
        \codefont{SSLContext} & 26 & 26 \\
        \hline
        \codefont{TrustManager} &248 &248 \\
        \bottomrule
        \textbf{Total}  &\vcount & \vcount\\
        \bottomrule
    \end{tabular}
    \vspace{-1.5em}
    \end{table}

By applying \tool to the second data set mentioned in Section~\ref{sec:evaluateApply}, \tool reported \vcount real vulnerabilities in 100 open-source projects. About 43\% of these vulnerabilities (i.e., 422) are related to \codefont{SecureRandom} APIs; 25\% of vulnerabilities are about \codefont{TrustManager} API misuses; and 15\% of vulnerabilities are relevant to \codefont{MessageDigest}. To assess the quality of \tool's fix suggestions, we manually checked 100 fixes suggested by \tool. To ensure the representativeness of our manual inspection results, we selected the 100 fixes to cover at least one vulnerability for each security class API. Based on the our inspection, the 100 fixes are all correctly customized for program contexts. It means that \tool can properly suggest fixes. 

To further investigate how developers assess the quality of \tool's outputs, we picked \prCount vulnerable Java classes reported by \tool and filed \prCount pull requests (PRs) in the project repositories. In each PR, we (1) described the vulnerable code, (2) suggested the fix(es), (3) explained the security implication of each vulnerability, and (4) asked for developers' opinions on the program issues and solutions. Up till now, we have received developers' feedback for 25 PRs. As shown in Table~\ref{tab:feedback}, developers confessed the revealed vulnerabilities in 14 PRs, needed us to provide more clarification information for 3 PRs, and believed the reports to be security-irrelevant for 8 PRs. More importantly, for 8 of the 14 PRs, developers took actions to either fix the vulnerabilities, disable the insecure implementation, or at least leave comments to warn people of the insecure code. 

\begin{table}
\caption{Developers' opinions on \tool's outputs}\label{tab:feedback}
\footnotesize
\vspace{-1.2em}
\begin{tabular}{lr} \toprule
\textbf{Feedback} & \textbf{\# of Pull Requests} \\ \toprule
Confess vulnerabilities & 14  \\
Need more clarification & 3 \\
Believe to be irrelevant & 8 \\ \hline
Try to resolve the security issues & 6 \\ \bottomrule
\end{tabular}
\vspace{-1.5em}
\end{table}

Interestingly, some developers rejected our repair suggestions for three major reasons. First, several developers would like to see real attacks to exploit those vulnerabilities (e.g., \codefont{TrustManager} trusts all). Particularly, two developers emphasized that the vulnerabilities (e.g., \codefont{MessageDigest.getInstance("MD5")} exist in test cases, which implies that these vulnerabilities cannot lead to real attacks. 
Second, some developers are reluctant to introduce API breaking changes by removing the insecure functionalities from their software ($P$), because they are afraid that the client projects based on their software ($CP$) can be negatively influenced (e.g., getting compilation errors). Therefore, these developers rely on the wisdom of $CP$ programmers to (1) identify the insecure options provided by $P$ and (2) extend $P$ to implement stronger security protection as needed. 

Third, certain developers believe that some reported API misuses are not necessarily insecure. For instance, one developer explained that his Java method \codefont{computeSha256(File)} invokes  \codefont{MessageDigest.getInst\-ance("SHA-256")} just to generate a message digest for any given file. This functionality has nothing to do with security. 

\vspace{0.5em}
\noindent\begin{tabular}{|p{8cm}|}
	\hline
	\textbf{Finding 3:} \emph{\tool revealed \vcount vulnerabilities in 100 open-source projects. By filing \prCount PRs based on some of \tool's outputs, we identified a mixture of developers' opinions on the revealed vulnerabilities and suggested fixes.
	}
\\
	\hline
\end{tabular}
\vspace{1em}

%% file: Related.tex
\section{Related Work}
The related work includes automatic program repair,  detection of security API misuses, and example-based program transformation.  

\vspace{-0.5em}
\subsection{Automatic Program Repair (APR)}
Tools were proposed to generate candidate patches for certain bugs, and to automatically check patch correctness using compilation and testing~\cite{Mechtaev2016,Long2016:Prophet,Qi2014,DeMarco2014,Kim2013:PAR,LeGoues12:gp}. Such approaches usually make two assumptions. First, software bugs can trigger some test failure(s). Second, automatic fault localization techniques~\cite{Jones2005,SP:Lo2010,Dao:2017} can be adopted to locate bugs in source code based on the execution coverage of passed and failed tests. With such assumptions, GenProg~\cite{LeGoues12:gp} and RSRepair~\cite{Qi2014} create  candidate patches by replicating, mutating, or deleting code randomly from the existing buggy program. 
To further improve the quality of created patches, PAR~\cite{Kim2013:PAR} and Prophet~\cite{Long2016:Prophet} prioritize patch generation based on the frequently applied bug fixes by developers. Each patch generated by current APR tools can only edit one or two lines of code (e.g., changing one or two numeric values or adding an \codefont{if}-condition check).  

Existing APR tools are not applicable to vulnerability detection and repair for three reasons. First, their assumptions do not hold. Vulnerable code is functionally correct and rarely fails any test case,  so APR tools cannot rely on test failures to reveal vulnerabilities or assess applied fixes. Second, given a security API (e.g., \codefont{MessageDigest.getInstance(...)}), the secure parameter options can be totally different from insecure ones (e.g., \codefont{SHA-512} vs.~\codefont{MD1}). Therefore, it is almost impossible for APR tools to suggest secure options based on the existing codebases with solely insecure options. Third, some security patches require for significant code modification (e.g., inserting more than 10 lines of code), but APR tools are unable to generate such complex edits.  

\vspace{-0.5em}
\subsection{Detecting of Security API Misuses}
Researchers built various tools to detect security vulnerabilities related to API misuses~\cite{fahl2012eve,egele2013empirical,he2015vetting,fischer2017stack,kruger2018crysl,Fischer2019,Rahaman2019}.
Specifically, most tools conduct static program analysis to check programs against hardcoded templates of vulerabilities~\cite{fahl2012eve,egele2013empirical,he2015vetting,Rahaman2019}. For instance, 
MalloDroid scans the decompiled code of Android apps to a) extract the networking API calls and valid HTTP(S) URLs, b) check the validity of the SSL certificates of all extracted HTTPS hosts, and c) identify apps that invoke APIs to customize SSL usage~\cite{fahl2012eve}. Some other tools adopt machine learning (ML)-based approaches to identify vulnerabilities~\cite{fischer2017stack,Fischer2019}. For instance, Fischer et al.~\cite{fischer2017stack} transformed secure/insecure code corpus into numeric vectors, and used support vector machine (SVN) to train a binary-class classifier based on those vectors. Next, given arbitrary code $C$, the classifier predicts whether C is insecure. Additionally, CrySL is a domain-specific language (DSL) that developers can use to describe vulnerable usage of security APIs~\cite{kruger2018crysl}. 


However, it is not easy to extend the above-mentioned tools to detect new vulnerabilities. For instance, the tools based on hardcoded rules often require users to modify the tool implementation; ML-based tools may require for a large amount of labeled secure and insecure code samples for classifier retraining; DSL-based tools require users to learn new languages for pattern prescription, while the learning curve may be high for some security researchers or developers. In comparison, \tool has great extensibility because it only requires users to demonstrate vulnerabilities and related repairs with a small number of Java code examples. By inferring patterns from those examples, \tool gains the capability of detecting new vulnerabilities and proposing related fixing suggestions. 


\vspace{-0.5em}
\subsection{Example-Based Program Transformation}

Based on the insight that developers modify similar code in similar ways, researchers developed tools to infer program transformations from exemplar edits and to manipulate code or suggest changes accordingly~\cite{Miller2001,Toomim2004,Duala-Ekoko2007,Meng2013:lase,ma2017vurle,rolim2017learning,An2018,Xu2019}. For instance, while users interactively edit code in one program context, simultaneous text editing approaches apply identical textual edits in other preselected program contexts at the same time~\cite{Miller2001,Toomim2004,Duala-Ekoko2007}. Given one or multiple code change examples, LASE~\cite{Meng2013:lase} and REFAZER~\cite{rolim2017learning} infer a program transformation from the examples; they then use the transformation to locate similar code to edit, and apply customized transformations to those locations. TINFERER generalizes language translation rules from exemplar Java code and the equivalent implementation by Swift; it then applies the Java-to-Swift rules to translate any given Java program into Swift~\cite{An2018}. The most closely related work to \tool is VuRLE~\cite{ma2017vurle}, which also detects and repairs vulnerabilities based on <insecure, secure> code examples. 

However, simultaneous editing cannot identify edit locations or apply customized edits. The other approaches only adopt (1) token-based or AST-based matching and/or (2) simple intra-procedural analysis to identify edit locations; they are insufficient for two reasons. First, token-based or AST-based statement matching is inflexible. When an edit location has statements formatted differently from the exemplar code (compare Figure~\ref{fig:example}($I$) with Listing~\ref{lst:p}), current tools can miss the location. Second, intra-procedural analysis only finds code matches in single methods; it does not go beyond method boundaries or reveal any edit location that involves code from multiple methods. In comparison, \tool conducts inter-procedural analysis and program slicing to overcome the limitations. Actually, 
we also contacted the authors of VuRLE to ask for their programs and data, but did not get the materials to empirically compare \tool with VuRLE.

%% file: Threats.tex
\section{Threats to validity}
\label{sec:threats}
 All inferred patterns and detected vulnerabilities are limited to our experiment data sets. The observations may not generalize well to close-source projects. In the future, we plan to include more <insecure, secure> examples or even close-source projects into our evaluation, so that our findings are more representative. 

We evaluated the recall rates of \tool and CryptoGuard based on a set of \vulCount vulnerabilities manually found in 10 open-source projects. This ground truth set is subject to human bias, and our data set constructor accidentally overlooked some actual vulnerabilities in those projects. Such incomplete ground truth can influence recall evaluation. In the future, we plan to recruit more security experts to simultaneously inspect the same codebases, so that the built ground truth set is more complete. 

In some repair suggestions provided by \tool, there are placeholders that we need developers to further customize (see ``\codefont{//Please change `example.com' as needed}'' in Figure~\ref{fig:example3}). Such placeholders should be filled based on developers' unique software environments (e.g., host name settings), or sometimes even require developers to do extra configurations outside the codebase (e.g., installing SSL certificates). Therefore, the current repair suggestions by \tool may sometimes seem vague and abstract. In the future, we plan to provide more detailed suggestions and explore interactive approaches that guide developers to apply complete repairs step-by-step. 


%% file: Conclusion.tex
\section{Conclusion}
Security is important for software quality assurance. 
To help developers better protect their software based on the state-of-the-art security research, in this paper, we present \tool---a new approach that takes in  <insecure, secure> code examples, generalizes vulnerability-repair patterns from examples, and applies those patterns for vulnerability detection and repair suggestion. Compared with prior work, \tool lowers the technical barriers for security experts to illustrate their domain knowledge, and concretizes security expertise as customized code edits for software practitioners.
In this way, \tool intends to bridge the gap between security research and software development.  

Based on our evaluation results, \tool is capable of inferring various patterns from diverse code examples; it detects real vulnerabilities and suggests fixes in open-source projects with high accuracy. More importantly, we filed \prCount pull requests (PRs) to seek for developers' opinions on the revealed vulnerabilities and fix suggestions by \tool. Interestingly, responders of \vulConfirmed PRs agreed upon the revealed vulnerabilities; for 8 of these PRs, responders took actions to mitigate or eliminate those security issues. Meanwhile, responders of another eight PRs disliked \tool's outputs due to (1) the fear of breaking their software specifications when applying security patches, (2) expectation for real exploits of those vulnerabilities, (3) belief in software users to consciously adopt vulnerable code for good reasons (e.g., quick prototyping), and (4) disagreement on certain security API misuses. Our findings show interesting opinion contrasts between security research and software practices. In the future, we will investigate approaches to automate security attacks based on revealed vulnerabilities.

